\documentclass[twocolumn,showpacs,preprintnumbers,amsmath,amssymb]{revtex4}

\usepackage{graphicx}
\usepackage{dcolumn}
\usepackage{bm}

\begin{document}


\title{Vortex Lattice Structures of a Bose-Einstein Condensate in a Rotating Triangular
Lattice Potential}

\author{T.~Sato$^1$}
\author{T.~Ishiyama$^2$}
\author{T.~Nikuni$^2$}

\affiliation{
$^1$Institute for Solid State Physics, The University of Tokyo, 5-1-5 Kashiwanoha, Kashiwa, Chiba 277-8581, Japan\\
$^2$Tokyo University of Science, 1-3 Kagurazaka, Shinjuku-ku, Tokyo, 162-9601, Japan}

\date{\today}

\begin{abstract}
We study the vortex pinning effect in a Bose-Einstein Condensate in the presence of a rotating lattice potential
by numerically solving the time-dependent Gross-Pitaevskii equation.
We consider a triangular lattice potential created by blue-detuned laser beams.
By rotating the lattice potential, we observe a transition from the Abrikosov
vortex lattice to the pinned vortex lattice.
We investigate the transition of the vortex lattice structure by changing conditions such as angular
velocity, strength, and lattice constant of a rotating lattice potential.
Our simulation results clearly show that the lattice potential has a 
strong vortex pinning effect when the vortex density coincides with the density of the pinning points.

\end{abstract}

\maketitle

\section{\label{sec:level1}Introduction}
Quantized vortices are one of the most characteristic manifestations of superfluidity associated with a 
Bose-Einstein condensate (BEC) in atomic gases~\cite{AD}.
Formations of triangular Abrikosov vortex lattice in Bose condensates have been clearly observed
by rotating anisotropic trap potentials~\cite{KM,FC,JA}.
Microscopic mechanism of the vortex lattice formation has been extensively studied both analytically and
numerically using the Gross-Pitaevskii equation for the condensate wavefunction~\cite{DF,MT,MU,KK,NS,KKA,AF,TS}.

Another interesting development in ultracold atomic gases is creating periodic potential using optical lattices~\cite{MM}.
The studies of a BEC in optical lattices have found a lot of interesting phenomenon, such as
the transition from the superfluid to Mott insulator phase~\cite{Gr}.

More recently, combining the above two systems, i.e., rotating BEC and optical lattices has attracted
growing attention.
In particular, the vortex phase diagrams of a BEC in a rotating optical lattice potential have been theoretically
studied, since one expects structural phase transition of vortex lattice structures due to vortex pinning~\cite{RB,JR,HPP,HP}.
In the condensed matter systems such as superconductors, the vortex pinning due to impurities
and defects in solids have been extensively investigated~\cite{PP,PPP}.
In the atomic condensates, rotating optical lattices have been experimentally realized at JILA group, making use of
a laser beam passing through a rotating mask~\cite{ST}.
This experiment observed the structural phase transition of vortex lattice structures in rotating Bose condensated gases
due to vortex pinning by the laser beam.
In this system, one can control the pinning parameters by changing conditions such as angular velocity, strength,
and lattice constant of rotating optical lattices. 
This allows one to investigate vortex pinning effect in a quantitative manner..

In this paper, stimulated by the JILA experiment, we study the vortex pinning effect by numerically solving the time-dependent
Gross-Pitaevskii equation.
We observe a structural phase transition of vortex lattice structures of a BEC in 
a rotating triangular lattice potential created by blue-detuned laser beams.

\section{\label{sec:level1}Formulation}
We consider a Bose-condensated gas trapped in a harmonic potential and a rotating periodic potential
with an angular velocity $\Omega$.
The dynamics of the condensate is described by the time-dependent Gross-Pitaevskii equation.
Assuming a pancake-shaped trap, we use a two-dimensional Gross-Pitaevskii equation.
In the rotating frame, the Gross-Pitaevskii equation is given by (for a detailed derivation,
see for example Ref.~\cite{MU})
\begin{eqnarray}
(i-\gamma)\hbar\frac{\partial \psi(\mathbf r,t)}{\partial t}=\bigg[-\frac{\hbar^2}{2m}\nabla^2+V_{\rm ext}(\mathbf r)+g{\vert \psi(\mathbf r,t) \vert}^2 \nonumber\\
                                                            -\Omega L_z\bigg]\psi(\mathbf r,t).
\label{eq:GP}
\end{eqnarray}
Here, $V_{\rm ext}(\mathbf r)=m\omega_{0}^{2}(x^2+y^2)/2+V_{\rm lattice}(\mathbf r)$
describes the total external potential, $L_z=-i\hbar (x \partial/\partial y- y\partial/\partial x)$
denotes the $z$ component of the angular momentum operator, $g=4 \pi \hbar^2 a_{\rm s}/m$ is the strength of
interaction with $a_{\rm s}$ being the $s$-wave scattering length, and $ \gamma$ is the
phenomenological dissipative parameter~\cite{SC,MO,EZ,TN,DS}.
The lattice potential created by blue-detuned laser beams arranged in the lattice
geometry is expressed as
\begin{eqnarray}
V_{\rm lattice}(\mathbf r) = \sum_{n_1,n_2}V_0\exp
\left[-\frac{\vert \mathbf r-  \mathbf r_{n_1,n_2}\vert^2}{(\sigma/2)^2}\right],
\label{eq:one}\end{eqnarray}
where $ \mathbf r_{n_{1},n_{2}}=n_{1}\mathbf a_{1}+n_{2}\mathbf a_{2}$ describes lattice pinning points and 
$V_0$ describes the strength of the laser beam.
We consider the triangular lattice geometry with two lattice unit vectors given by $\mathbf a_{1}=a(1,0)$,
$ \mathbf a_{2}=a(-1/2,\sqrt{3}/2)$, where $a$ is the lattice constant.
Throughout this paper, we scale the length and energy by $a_{\rm ho}=\sqrt{\hbar/m\omega_0}$ and $\hbar\omega_0$.
We set the dimensionless interaction strength as $C=4 \pi a_{s}N/R=1000$, where $N$ is the total particle number
and $R$ is size of the condensate along the $z$-axis direction~\cite{MU}, and the width of the laser beam as $ \sigma/a_{\rm ho}=0.65$.
In our parameter set, the healing length is $ \xi/a_{\rm ho}=0.12$.
We numerically solve the Gross-Pitaevskii equation in Eq.~(\ref{eq:GP})
using Fast-Fourier-Transform (FFT) technique.

We first determine the ground-state condensate wavefunction without rotation by setting $\Omega=0$ in Eq.~(\ref{eq:GP}). 
Starting with this wavefunction as a initial state, we numerically evolve the Gross-Pitaevskii equation with a rotating
lattice potential with a fixed angular velocity $\Omega$, until the system relaxes into equilibrium.
We investigate the vortex lattice structure from the equilibrium condensate density profile 
$n(\mathbf r)=\vert \psi(\mathbf r)\vert^2 $.
In order to quantify the vortex lattice structure, we calculate the density structure factor defined by
\begin{eqnarray}
S( \mathbf k)  = \int d \mathbf r n(\mathbf r) e^{-i \mathbf k \cdot \mathbf r}.
\label{eq:Sk}
\end{eqnarray}
The structure factor given in Eq.~(\ref{eq:Sk}) contains information about the periodicity of the condensate density.
For triangular lattice geometry,$S({\bf K})$ exhibits periodic peaks of the regular hexagonal geometry.
In order to distinguish between the Abrikosov lattice and the pinned vortex lattice, we calculate the peak intensity
of the structure factor $|S({\bf K})|$ at the lattice pinning point, namely,
\ $\mathbf K_{1}=2\pi/a(1,1/\sqrt{3})$, $ \mathbf K_{2}=2\pi/a(0,2/\sqrt{3})$.
In addition to the structure factor, we also calculate the lattice potential energy
defined by
\begin{eqnarray} E_{\rm lattice} = 
\int {d\mathbf r \psi^*(\mathbf r)V_{\rm lattice}(\mathbf r) \psi(\mathbf r)},\label{eq:one}
\end{eqnarray}
\begin{figure}
\centerline{\includegraphics[height=1.3in]{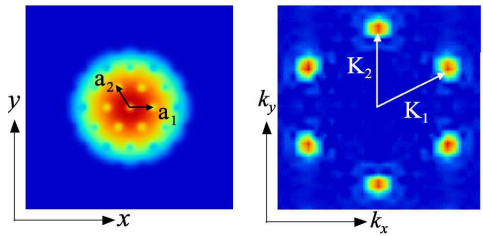}}
\caption{Density profile (left) and Structure factor profile (right) of a BEC in a triangular lattice potential without
rotation by setting $\Omega=0$.
The lattice potential geometry is triangular lattice at $a/a_{\rm ho}=2.2$ and $V_0/\hbar\omega_0=5.0$.}  
\label{fig:tau2}
\end{figure}
which can also be used to quantify the transition of the vortex structure.
As we see below, the vortex pinning is signified by a marked decrease of $E_{\rm lattice}$.

\section{\label{sec:level1}Simulation Results}

\begin{figure}
\centerline{\includegraphics[height=4.1in]{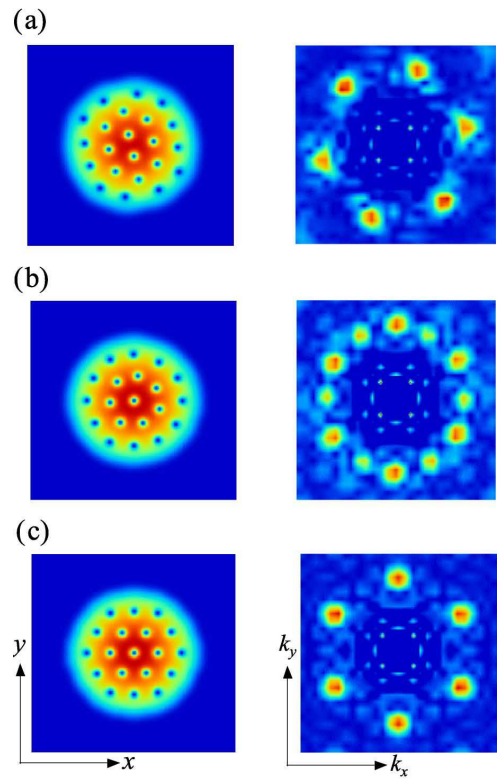}}
\caption{Density profiles~(left) and Structure factor profiles~(right) in equilibrium state after rotating condensates
with a fixed angular velocity $ \Omega/\omega_0=0.70$ at $a/a_{\rm ho}=2.2$.
The figures show vortex lattice structures for different values of the strength of a triangular lattice potential;
(a) $V_0/\hbar\omega_0=0.1$, (b) $V_0/\hbar\omega_0=0.3$, (c) $V_0/\hbar\omega_0=0.5$.}  
\label{fig:tau2}
\end{figure}
In this section, we show our numerical simulation results for vortex lattice structures of a BEC in a rotating triangular
lattice potential.
Fig.~1 shows the equilibrium condensate density profile and the density structure factor profile in the presence of
a triangular lattice potential without rotation.

Fixing the lattice constant with $a/a_{\rm ho}=2.2$, we investigate the transition of the vortex lattice structure
by changing the strength $V_0$ for various angular velocities $ \Omega$.
Fig.~2 shows the density profiles and the structure factor profiles in equilibrium state after rotating condensates
with a fixed angular velocity $ \Omega/\omega_0=0.70$.
One can see that for weak lattice potential $V_0/\hbar\omega_0=0.1$ (Fig.~2 (a)), vortices form the Abrikosov lattice,
which is incommensurate with the triangular lattice potential.
By slightly increasing the lattice strength as $V_0/\hbar\omega_0=0.3$ (Fig.~2 (b)), vortices start to being
partially pinned by the triangular lattice potential.
For strong lattice potential $V_0/\hbar\omega_0=0.5$ (Fig.~2 (c)), all vortices are pinned by the triangular lattice potential.
We thus observed a transition of the vortex lattice structure from the Abrikosov lattice to the pinned lattice.

In Fig.~3, we plot $|S({\bf K})|$ and $E_{\rm lattice}$ against the lattice strength $V_0$.
As shown in Fig.~3 (a), for $\Omega/\omega_0=0.70$, the lattice potential energy $E_{\rm lattice}$ decreases gradually,
which indicates that vortices are partially pinned, and reaches constant when all vortices are pinned for $V_0/\hbar\omega_0\ge0.4$.
Correspondingly the structure factor $|S({\bf K})|$ increases gradually and finally becomes constant when all vortices are pinned.
From these results, together with directly looking at the condensate density profile, we conclude that the structural
phase transition occurs at $V_0/\hbar\omega_0=0.4$,
which we define as the strength for the structural phase transition $V_{\rm c}/\hbar\omega_0$. 
\begin{figure}
\begin{minipage}{0.5\linewidth}
\centering
\centerline{\includegraphics[height=2in]{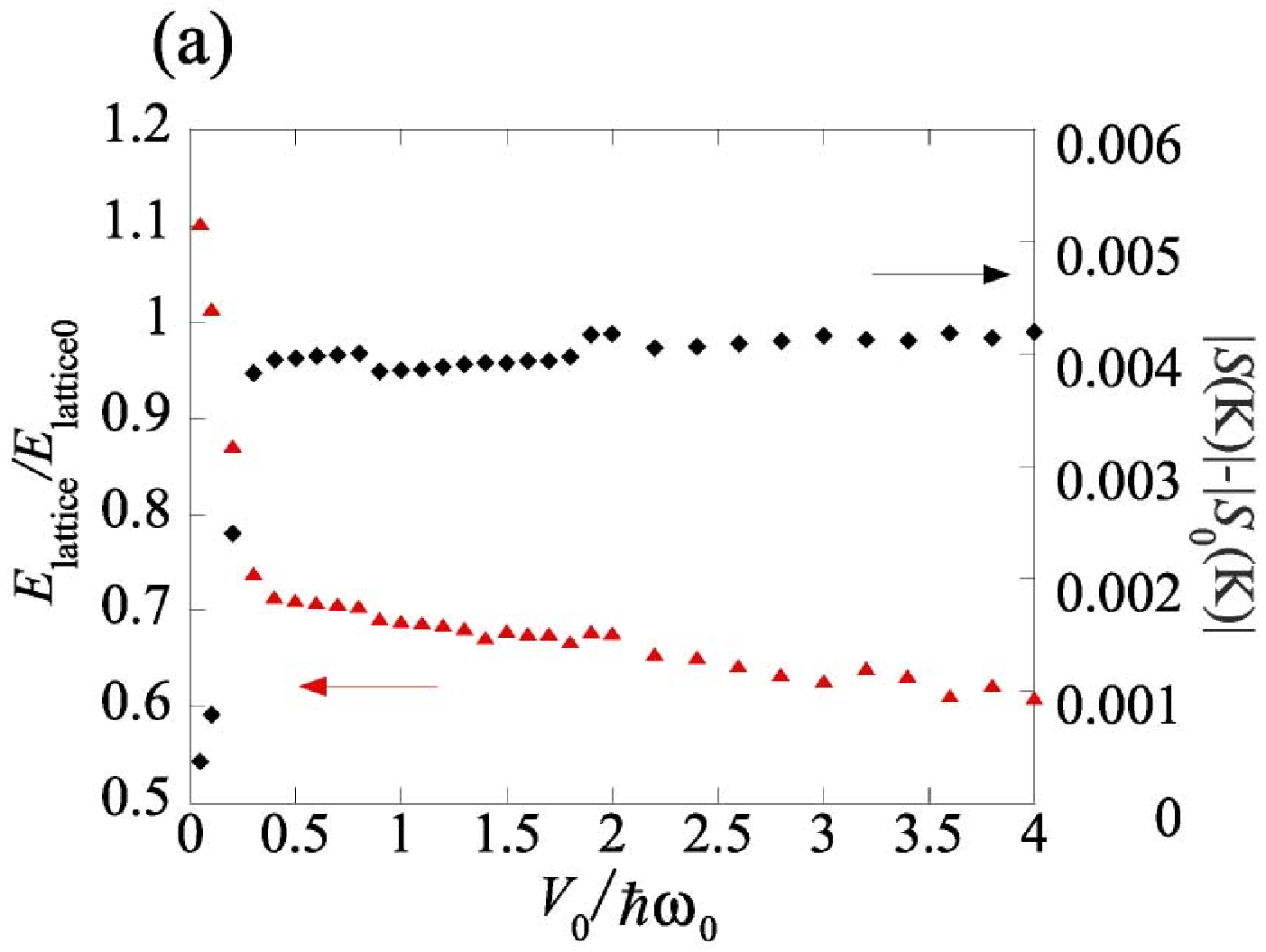}}
\end{minipage}
\begin{minipage}{0.5\linewidth}
\centering
\centerline{\includegraphics[height=2in]{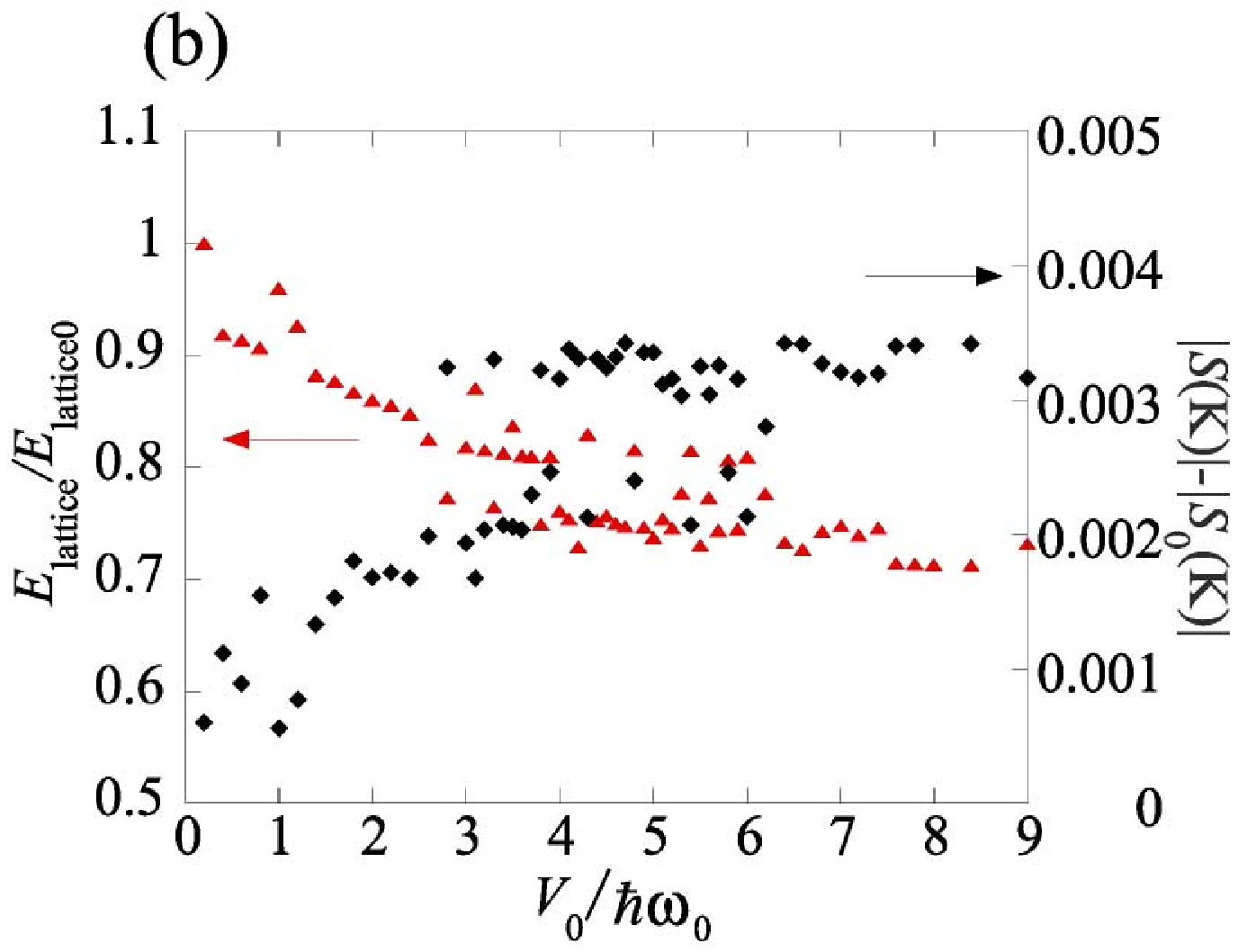}}
\end{minipage}
\caption{Lattice potential energy ($ \blacktriangle $) and peak intensity of the structure factor at the lattice
pinning point ($ \blacklozenge $) with $a/a_{\rm ho}=2.2$ corresponding to the angular velocity;
(a) $\Omega/\omega_0=0.70$, (b) $\Omega/\omega_0=0.55$.
Here $E_{\rm lattice0}$ and $|S_0({\bf K})|$ is the lattice energy and the peak intensity of the structure factor
at the lattice pinning point of the ground state ($\Omega=0$) for each lattice strength.}  
\label{fig:tau2}
\end{figure}
Fig.~3 (b) shows the analogous result for $ \Omega/\omega_0=0.55$.
In this case, there is an intermediate domain where the Abrikosov lattice and the pinned lattice coexist.
In this domain, the vortex lattice structure cannot be categorically determined because of the competition between
the vortex-vortex interaction and the lattice pinning effect.

\begin{figure}
\begin{minipage}{0.5\linewidth}
\centering
\centerline{\includegraphics[height=2in]{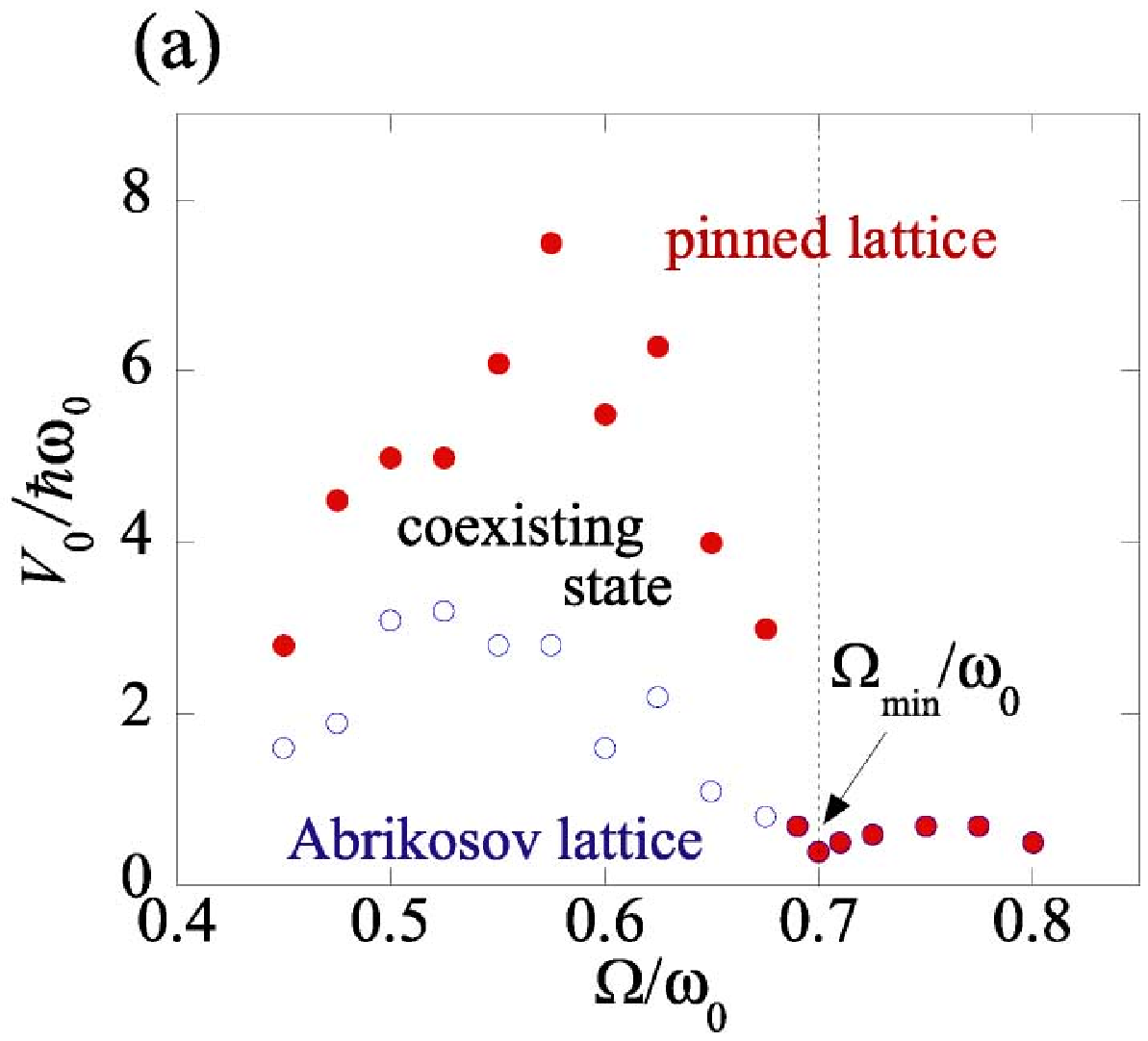}}
\end{minipage}
\begin{minipage}{0.5\linewidth}
\centering
\centerline{\includegraphics[height=2in]{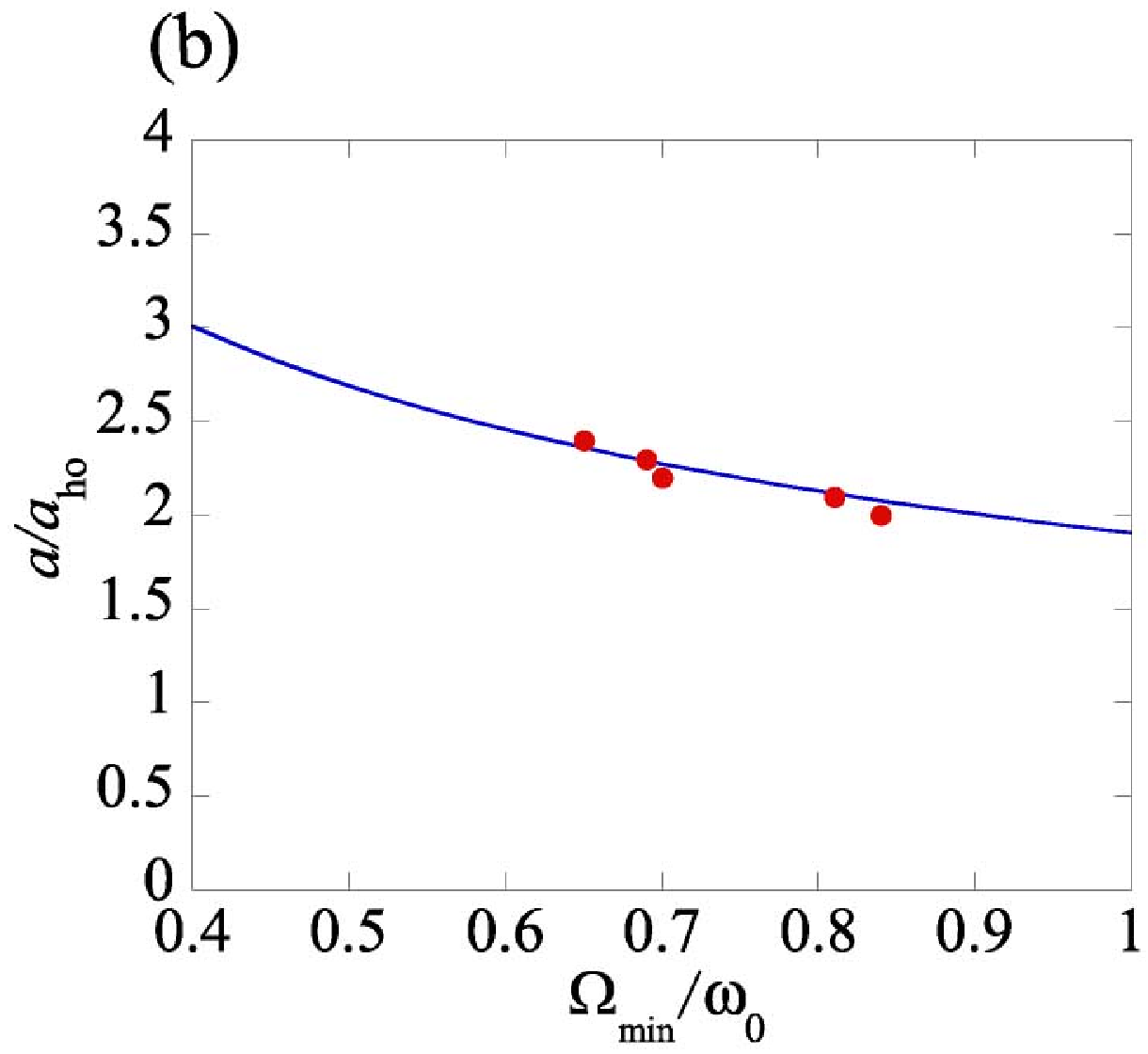}}
\end{minipage}
\caption{(a) Phase diagrams for vortex lattice structures for $a/a_{\rm ho}=2.2$.
(b) Lattice constant $a$ against $ \Omega_{\rm min}$ giving the minimum pinning strength.
The solid line represents Eq.~(5).}  
\label{fig:tau2}
\end{figure}

From these results, we map out the phase diagrams of vortex lattice structures against
$ \Omega$ and $ V_0$, as we plot in Fig.~4~(a).
The lower domain is the Abrikosov lattice domain, while the upper domain is the pinned lattice domain.
The intermediate domain represents coexisting state of the Abrikosov lattice and the pinned lattice.
Looking at the phase boundary of the pinned lattice domain in Fig.~4~(a) as 
$V_{\rm c}=V_{\rm c}(\Omega)$, we find that $ V_{\rm c}$ takes a minimum value as a function of $ \Omega$,
which we define as the minimum pinning strength
$V_{\rm c, min}=V_{\rm c}(\Omega=\Omega_{\rm min})$. 
From Fig.~4~(a) for $a/a_{\rm ho}=2.2$, 
we find that $V_{\rm c, min}/\hbar\omega_0 \thickapprox0.4$ and $ \Omega_{\rm min}/\omega_0 \thickapprox 0.70$.

The dependence of minimum pinning lattice strength $V_{\rm c, min}$ on the lattice constant $a$ can be understood as follows.
When vortices form the triangular lattice and undergo rigid rotation with angular velocity $ \Omega$,
the lattice constant is expressed as a function of angular velocity by
\begin{eqnarray}
\frac{a_{\rm v}(\Omega)}{a_{\rm ho}}=\sqrt{\frac{2}{\sqrt{3}} \frac{2\pi}{2\Omega/\omega_0}}.
\label{eq:a_omega-tr}
\end{eqnarray}

In Fig.~4 (b), we plot lattice constant $a$ against $ \Omega_{\rm min}$.
We find that the formula in Eq.~(\ref{eq:a_omega-tr}) well fits the numerical data. 
This means that the pinning strength $V_{\rm c}$ takes minimum value when $a_{\rm v}(\Omega)$
matches to the lattice constant $a$. 
When this ``matching relation" is satisfied, weak lattice potential has a stronger pinning effect than
vortex-vortex interaction, which leads to a sharp transition of the vortex lattice structure. 

\begin{figure}
\begin{minipage}{0.5\linewidth}
\centering
\centerline{\includegraphics[height=6.2in]{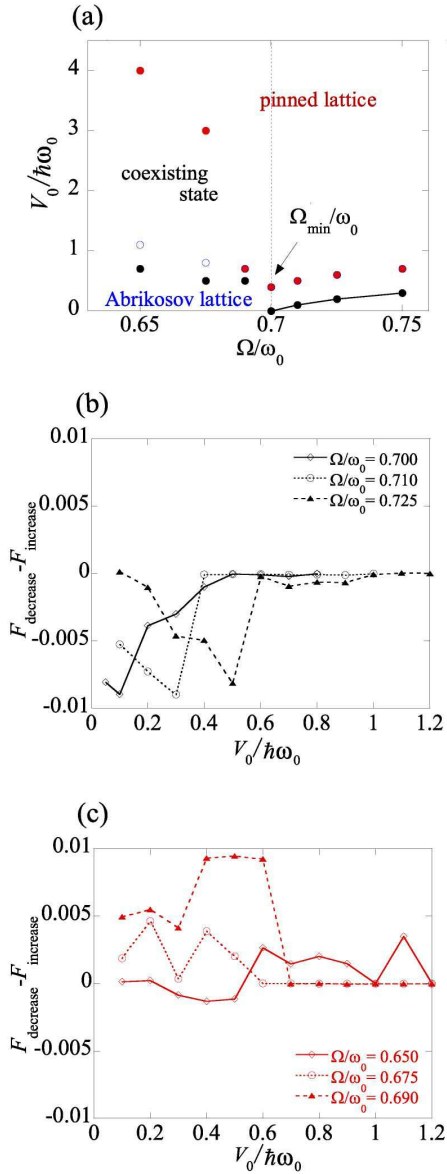}}
\end{minipage}
\caption{
(a) Phase diagrams for vortex lattice structures for $a/a_{\rm ho}=2.2$ by increasing $V_0$ or
decreasing $V_0$ from pinned lattice domain ($\blacklozenge$).
At $\Omega \geq \Omega_{\rm min}$, phase boundary of actually the pinned lattice phase is solid line.
(b) Total free energy $F$ with $a/a_{\rm ho}=2.2$ at $\Omega/\omega_0=0.70$ ($ \Diamond $),
$\Omega/\omega_0=0.71$ ($\bigcirc$) and $\Omega/\omega_0=0.725$ ($ \blacktriangle $).
Here $F_{\rm increase}$ is in the case of increase of $V_0$, while 
$F_{\rm decrease}$ is in the case of decrease of $V_0$ from pinned lattice domain.
(c) Total free energy $F$ with $a/a_{\rm ho}=2.2$ at $\Omega/\omega_0=0.65$ ($ \Diamond $),
$\Omega/\omega_0=0.675$ ($\bigcirc$) and $\Omega/\omega_0=0.69$ ($ \blacktriangle $).}  
\label{fig:tau2}
\end{figure}

In the case when this ``matching relation" is satisfied, one may presume that the vortex lattice array match 
the lattice potential array without vortex pinning effect.
However, from Fig.~4 (a), we find that minimum pinning lattice strength $V_{\rm c, min}$ does not fall to zero, but takes a finite value.
Actually in obtaining the phase diagram in Fiq.~4 (a), we solved the Gross-Pitaevski equation starting
with the initial ground state wavefunction without vortices.
As evolving the Gross-Pitaevski equation, vortices are nucleated, forming vortex lattices.
In this dynamical process, it may be possible that vortices relax into metastable configurations, and this may
be the reason why $V_{\rm c, min}$ takes a finite value.
In order to investigate this in more detail, we solve the Gross-Pitaevski equation starting with a perfectly pinned
vortex lattice at large enough $V_0$.
We then slowly decrease $V_0$.
In Fig.~5 (a), we plot the phase boundary of the pinned lattice phase obtained in this way, together with phase boundaries previously
shown in Fiq.~4 (a).
We find that at $\Omega=\Omega_{\rm min}$, the perfectly pinned vortex lattices is stable down to infinitesimally small
lattice potential $V_0 \to 0$.
In order to see whether the pinned state obtained here has the lower energy than the tilted Abrikosov lattice obtained
in the previous calculation, we calculate the total free energy
\begin{eqnarray} 
F=\Big < \Big[-\frac{\hbar^2}{2m}\nabla^2+V_{\rm ext}(\mathbf r)+g{\vert \psi(\mathbf r,t) \vert}^2 \nonumber\\
-\Omega L_z \Big ] \psi(\mathbf r,t) \Big >.
\label{eq:F}
\end{eqnarray}
In Figs.~5 (b) and (c), we plot the free energy $F$ for $a/a_{ho}=2.2$ against $V_0$.
Here, $F_{\rm increase}$ is the free energy obtained by solving the Gross-Pitaevski equation starting with non-vortex initial state
and increasing $V_0$,
while $F_{\rm decrease}$ is the free energy obtained by decreasing $V_0$ from pinned lattice domain.
We make the separate plots for $\Omega \geq \Omega_{\rm min}$ and $\Omega < \Omega_{\rm min}$
in Figs.~5 (b) and (c), respectively.
One can see from Fig.~5 (b) that, at $\Omega=\Omega_{\rm min}$, $F_{\rm decrease}<F_{\rm increase}$
holds down to $V_0 \to 0$.
This means that the Abrikosov lattice found for $0<V_0<V_{\rm c, min}$ in the phase diagram in Fig.~4 (a) is
actually metastable.
Similarly for $\Omega>\Omega_{\rm min}$, we find a finite region where the Abrikosov lattice is metastable.
However, for $\Omega<\Omega_{\rm min}$, we find $F_{\rm decrease}>F_{\rm increase}$ as shown in Fiq.~5 (c).
This means that the pinned vortex lattice does not have the lowest energy in the weak lattice domain.

From the above results, we found that property of vortex pinning is quite different for $\Omega \geq \Omega_{\rm min}$ 
and $\Omega<\Omega_{\rm min}$. 
For $\Omega \geq \Omega_{\rm min}$, the vortex lattice structure exhibits sharp transition.
In contrast, for $\Omega<\Omega_{\rm min}$, there is no sharp transition, but the vortex lattice structure
exhibits crossover through the intermediate coexisting region.

\section{\label{sec:level1}Summary and Discussion}
In summary, we have studied the vortex pinning effect by observing the structural phase transition of vortex lattice
structures of a Bose-Einstein condensate in a rotating triangular lattice potential. 
We observed the transition of the vortex lattice structure from the Abrikosov vortex lattice to the pinned lattice
where all vortices are pinned to lattice points.
The transition is determined depending on the competition between strength and density of the lattice
potential, and vortex density and vortex-vortex interaction.
From our simulation, we found that the lattice potential has a strong vortex pinning effect when $a_v(\Omega)$
matches to the lattice constant $a$ of the triangular lattice potential, which means that the vortex density coincides
the density of pinning points.
In the case when this ``matching relation" is satisfied, we observed that minimum pinning lattice strength
$V_{\rm c, min}$ takes a finite value.
By investigating the structural transition in more detail, we found that for $\Omega \geq \Omega_{\rm min}$, there are regions
where the Abrikosov lattice is metastable, while for $\Omega<\Omega_{\rm min}$, the pinned vortex lattice phase is
metastable state.
From the above results, we found that property of vortex pinning is quite different for $\Omega \geq \Omega_{\rm min}$ 
and $\Omega<\Omega_{\rm min}$. 
For $\Omega \geq \Omega_{\rm min}$, the vortex lattice structure exhibits sharp transition.
In contrast, for $\Omega<\Omega_{\rm min}$, there is no sharp transition, but the vortex lattice structure exhibits
crossover through the intermediate coexisting region.

In a separate paper, we will discuss the pinning effect of a rotating square lattice potential.
We will show that the property of vortex pinning is quite different from the triangular lattice
and more complicated for a rotating square lattice potential.

\section{\label{sec:level1}Acknowledgments}
We thank S. Konabe and S. Watabe for useful discussions and comments. 
We also thank N. Sasa for helpful comments on the numerical simulations. 
This research was supported by Academic Frontier Project (2005) of MEXT.


\begin{thebibliography}{10}
\bibitem{AD} N. G. Parker, B. Jackson, A. M. Martin, and C. S. Adams, condmat 07040146 (2007).
\bibitem{KM}K. W. Madison, F. Chevy, W. Wohlleben, and J. Dalibard, Phys. Rev. Lett {\bf 84}, 806 (2000).
\bibitem{FC}K. W. Madison, F. Chevy, V. Bretin, and J. Dalibard, Phys. Rev. Lett {\bf 86}, 4443 (2001).
\bibitem{JA}C. Raman, J. R. Abo-Shaeer, J. M. Vogels, K. Xu, and W. Ketterle, Phys. Rev. Lett {\bf 87}, 210402 (2001).
\bibitem{DF} David L. Feder and Charles W. Clark, Phys. Rev. Lett {\bf 87}, 190401 (2001).
\bibitem{MT} M. Tsubota,  K. Kasamatsu, and M. Ueda, Phys. Rev. A {\bf 65}, 023603 (2002)
\bibitem{MU} K. Kasamatsu, M. Tsubota, and M. Ueda, Phys. Rev. A {\bf 67}, 033610 (2003)
\bibitem{KK} K. Kasamatsu, M. Tsubota, and M. Ueda, Phys. Rev. A {\bf 66}, 053606 (2002)
\bibitem{NS} N. Sasa, M. Machida, and H. Matsumoto, J. Low Temp. Phys {\bf 138}, 617 (2005)
\bibitem{KKA} K. Kasamatsu, M. Machida, N. Sasa, and M. Tsubota, Phys. Rev. A {\bf 71}, 063616  (2005)
\bibitem{AF} Alexander L. Fetter, B. Jackson, and S. Stringari, Phys. Rev. A {\bf 71}, 013605 (2005)
\bibitem{TS} T. P. Simula, A. A. Penckwitt, and R. J. Ballagh, Phys. Rev. Lett {\bf 92}, 060401 (2004) 
\bibitem{MM} O. Morsch and M. Oberthaler, Rev. Mod. Phys {\bf 78}, 179 (2006)
\bibitem{Gr} M. Greiner, O. Mandel, T. Esslinger, T. W. H$\ddot{\rm a}$nsch, and I. Bloch, Nature {\bf 415}, 39 (2002).
\bibitem{RB} R. Bhat, L. D. Carr, and M. J. Holland, Phys. Rev. Lett {\bf 96}, 060405 (2006)
\bibitem{JR} J. W. Reijnders and R. A. Duine, Phys. Rev. Lett {\bf 93}, 060401 (2004).
\bibitem{HPP} H. Pu, L. O. Baksmaty, S. Yi, and N. P. Bigelow, Phys. Rev. Lett {\bf 94}, 190401 (2005)
\bibitem{HP} K. Kasamatsu and M. Tsubota, Phys. Rev. Lett {\bf 97}, 240404 (2006) 
\bibitem{PP} M. Baert, V. V. Metlushko, R. Jonckheere, V. V. Moshchalkov, and Y. Bruynseraede, Phys. Rev. Lett {\bf 74},
3269 (1995)
\bibitem{PPP} A. N. Grigorenko, S. J. Bending, M. J. Van Bael, M. Lange, V. V. Moshchalkov, H. Fangohr, and P. A. J. de Groot,
Phys. Rev. Lett {\bf 90}, 237001 (2003)
\bibitem{ST} S. Tung, V. Schweikhard, and E.A. Cornell, Phys. Rev. Lett {\bf 97},240402 (2006).
\bibitem{SC} S. Choi, S. A. Morgan, and K. Burnett, Phys. Rev. A {\bf 57}, 4057 (1998)
\bibitem{MO}M. -O. Mewes, M. R. Andrews, N. J. van Druten, D. M. Kurn, D. S. Durfee, C. G. Townsend, and W. Ketterle, Phys. Rev. Lett {\bf 77},
988 (1996)
\bibitem{EZ} B. Jackson and E. Zaremba, Phys. Rev. Lett {\bf 88}, 180402  (2002)
\bibitem{TN} E. Zaremba, T. Nikuni, and A. Griffin, J. Low Temp. Phys {\bf 116}, 277 (1999)
\bibitem{DS} D. S. Jin, M. R. Matthews, J. R. Ensher, C. E. Wieman, and E. A. Cornell, Phys. Rev. Lett {\bf 78}, 764 (1997).
\end{thebibliography}
\end{document}